\documentclass[twoside]{LCWS11}
\usepackage[latin1]{inputenc}
\usepackage[dvips]{graphicx,epsfig,color}
\usepackage{wrapfig,rotating}
\usepackage{amssymb,amsmath,array}
\usepackage{axodraw}
\usepackage{mathrsfs}
\usepackage{cite}
\newlength{\captsize} \let\captsize=\footnotesize
\newlength{\captwidth} 
\newlength{\beforetableskip} 
\newcommand{\capt}[1]{\begin{minipage}{\captwidth}
\let\normalsize=\captsize
\caption[#1] {#1}
\end{minipage}\\ \vspace{\beforetableskip}}

\newenvironment{Eqnarray}%
         {\arraycolsep 0.14em\begin{eqnarray}}{\end{eqnarray}}
\newcommand{\beqa}{\begin{Eqnarray}}
\newcommand{\eeqa}{\end{Eqnarray}}
\newcommand{\beq}{\begin{equation}}
\newcommand{\eeq}{\end{equation}}
\def\eq#1{eq.~(\ref{#1})}

\DeclareMathOperator{\Tr}{Tr}
\renewcommand{\Re}{{\rm Re}}
\renewcommand{\Im}{{\rm Im}}
\def\ifmath#1{\relax\ifmmode #1\else $#1$\fi}\def\lsim{{~\raise.15em\hbox{$<$}\kern-.85em\lower.35em\hbox{$\sim$}~}}
\def\gsim{{~\raise.15em\hbox{$>$}\kern-.85em\lower.35em\hbox{$\sim$}~}}
\def\ls#1{\ifmath{_{\lower1.5pt\hbox{$\scriptstyle #1$}}}}
\def\lsup#1{\ifmath{^{\lower1.5pt\hbox{$\scriptstyle #1$}}}}
\def\tanb{\tan\beta}
\def\phaa{\phantom{AA}}
\def\anti{\overline}
\def\wtil{\widetilde}
\def\mz{m_Z}
\def\mw{m_W}
\def\hpm{H^\pm}
\def\mhh{m_{H}}
\def\mhl{m_{h}}
\def\mha{m_{A}}
\def\mhpm{m_{H^\pm}}
\def\hsm{h\ls{\rm SM}}

\def\nn{\nonumber}
\def\half{\tfrac{1}{2}}
\def\quarter{\tfrac{1}{4}}
\def\eighth{\tfrac{1}{8}}
\def\abar{{\bar a}}
\def\bbar{{\bar b}}
\def\cbar{{\bar c}}
\def\dbar{{\bar d}}
\def\ebar{{\bar e}}
\def\fbar{{\bar f}}
\def\gbar{{\bar g}}
\def\hb{{\bar h}}

\def\ur{U_R}
\def\dr{D_R}
\def\mud{M_U}
\def\mdd{M_D}

\def\ddel{\!\!\mathrel{\raise1.5ex\hbox{$\leftrightarrow$\kern-.85em
\lower1.7ex\hbox{$\partial$}}}}

\begin{document}
\title{
A framework for precision 2HDM studies at the ILC and CLIC} 
\author{Howard E.~Haber
\vspace{.3cm}\\
Santa Cruz Institute for Particle Physics, University of California \\
1156 High Street, Santa Cruz, CA 95064 USA
}

\maketitle

\begin{abstract}
The precision measurements of Higgs boson observables will be critical in the
interpretation of the dynamics responsible for electroweak symmetry breaking.
The capabilities of the ILC and CLIC for precision Higgs studies are well documented.
In this talk, a theoretical framework is presented for interpreting phenomena that can arise
in the two-Higgs-doublet model (2HDM).  By imposing no constraints on the 
most general set of 2HDM parameters, the resulting formalism correctly identifies 
the physical 2HDM observables, which can be determined by model-independent experimental analyses.

\end{abstract}

\section{Introduction}

The capabilities of the ILC and CLIC for precision Higgs studies are well documented\cite{Djouadi:2007ik,Linssen:2012hp}.
Suppose that collider data suggests that electroweak symmetry breaking is a consequence of the
dynamics of a two-Higgs doublet model (2HDM).  To verify this hypothesis in detail will require
precision measurements of numerous Higgs boson observables.   However, to perform these 
measurements within the context of a general 2HDM is challenging, in part due to the numerous
parameters that govern the structure of the 2HDM Lagrangian.  

Indeed, a generic point of the 2HDM parameter space is incompatible
with current phenomenological constraints.  For example, without
imposing restrictions on the model, the 2HDM generically predicts
large Higgs-mediated tree-level flavor changing neutral currents
(FCNC) in
conflict with experimental observation\cite{Glashow:1976nt}.  It is
possible to remove the Higgs-mediated tree-level FCNCs
by imposing a simple discrete symmetry (a number of
different implementations, reviewed in ref.\cite{Branco:2011iw}, are
possible) or supersymmetry\cite{Fayet:1974fj,Gunion:1984yn}.
However, such symmetries are often broken, in which case the effective
2HDM below the symmetry-breaking scale would be a completely
general 2HDM.  Since there are many different ways to apply the
relevant symmetries, one does not know a priori which specific
realization of the 2HDM (if any) is the most likely to be realized in
nature.

Consequently, phenomenological 2HDM studies tended to be quite model
dependent.  However, it is preferable to let experiment decide which
realization of the 2HDM is relevant.  But to do so requires a
model-independent treatment of the 2HDM.  In such a treatment, the two
Higgs doublet fields are indistinguishable, in which case any two
linearly-independent combinations of the original two doublet fields
can be used to develop the 2HDM Lagrangian.  Physical observables
cannot depend on this choice of basis for the scalar fields.  By identifying the
relevant basis-independent (physical) observables, one could experimentally ``discover''
the presence of any approximate or exact symmetry of the 2HDM needed
for the phenomenological consistency of the model.

As a warmup, we consider some aspects of the precision Higgs program in the context of the
minimal supersymmetric standard model (MSSM) in Section 2. In Section 3, the 
basis-independent formalism for the 2HDM is developed and the physical observables (e.g., masses and couplings) are
identified.  The decoupling limit of the general 2HDM is introduced in Section 4, and its
significance is discussed.  Finally, conclusions and lessons for future work are outlined
in Section 5.

\section{What can we learn from precision MSSM Higgs studies?}

At the ILC\cite{precisionILC} and CLIC\cite{precisionCLIC}, it may be possible to measure the $h^0 b\bar b$ coupling to an accuracy
of a few percent or less, where $h^0$ is the lightest CP-even neutral Higgs boson.  A deviation from the prediction
of the Standard Model at some level is expected in models with an extended Higgs sector.  In this section, we consider the MSSM whose
Higgs sector is a special case of the 2HDM. 

\subsection{The wrong-Higgs interactions}

 In the MSSM, the tree-level Higgs--quark
Yukawa Lagrangian is supersymmetry-conserving and is given by:
$$
\mathcal{L}_{\rm yuk}^{\rm tree}=
-\epsilon_{ij}h_b H_d^i\psi_Q^j\psi_D
+\epsilon_{ij}h_t H_u^i\psi_Q^j\psi_U+{\rm h.c.}
$$
Two other possible dimension-four
gauge-invariant non-holomorphic Higgs-quark
interactions terms,  the so-called \textit{wrong-Higgs interactions}\cite{Carena:2002es,Haber:2007dj},
$$
H_u^{k*} \psi_{D} \psi_{Q}^k\qquad {\rm and} \qquad  H_d^{k*}\psi_U\psi_Q^k\,,
$$
are not supersymmetric (since the dimension-four supersymmetric Yukawa
interactions must be holomorphic), and hence
are absent from the tree-level MSSM Yukawa Lagrangian.

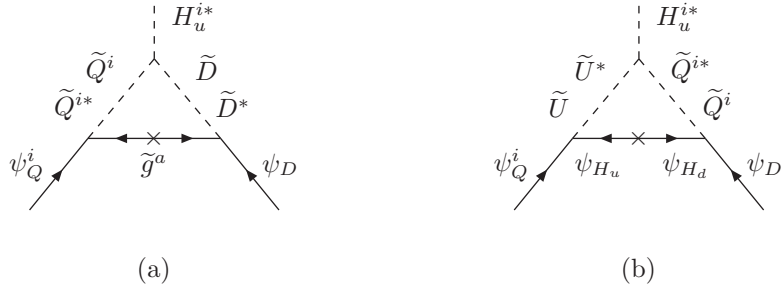
\begin{figure}[ht!]
\begin{center}
\begin{picture}(100,100)(40,0)
\DashLine(50,80)(50,60){3}
\ArrowLine(3,3)(25,30)
\ArrowLine(97,3)(75,30)
\ArrowLine(50,30)(75,30)
\ArrowLine(50,30)(25,30)
\DashLine(75,30)(50,60){3}
\DashLine(25,30)(50,60){3}
\Text(65,76)[]{$H_u^{i*}$}
\Text(20,43)[]{$\widetilde{Q}^{i*}$}
\Text(30,57)[]{$\widetilde{Q}^{i}$}
\Text(80,43)[]{$\widetilde{D}^{*}$}
\Text(70,57)[]{$\widetilde{D}$}
\Text(50,30)[]{$\times$}
\Text(50,20)[]{$\widetilde g^a$}
\Text(2,20)[]{$\psi_{Q}^i$}
\Text(98,20)[]{$\psi_D$}
\Text(50,-20)[]{(a)}
\end{picture}
\begin{picture}(100,100)(-40,0)
\DashLine(50,80)(50,60){3}
\ArrowLine(3,3)(25,30)
\ArrowLine(97,3)(75,30)
\ArrowLine(50,30)(75,30)
\ArrowLine(50,30)(25,30)
\DashLine(75,30)(50,60){3}
\DashLine(25,30)(50,60){3}
\Text(65,76)[]{$H_u^{i*}$}
\Text(20,43)[]{$\widetilde{U}$}
\Text(32,57)[]{$\widetilde{U}^*$}
\Text(80,43)[]{$\widetilde{Q}^{i}$}
\Text(70,57)[]{$\widetilde{Q}^{i*}$}
\Text(50,30)[]{$\times$}
\Text(35,20)[]{$\psi_{H_u}$}
\Text(68,20)[]{$\psi_{H_d}$}
\Text(2,20)[]{$\psi_{Q}^i$}
\Text(98,20)[]{$\psi_D$}
\Text(50,-20)[]{(b)}
\end{picture}
\end{center}
\vskip 0.1in
\caption{\small One-loop diagrams contributing to the
wrong-Higgs Yukawa effective operators\cite{Haber:2007dj}.  In (a), the cross ($\times$)
corresponds to a factor of the gluino mass $M_3$.  In (b), the
cross corresponds to a factor of the higgsino Majorana mass
parameter $\mu$. Field labels correspond to annihilation of
the corresponding particle at each vertex of the triangle.}\label{fig1}
\end{figure}

Nevertheless, the wrong-Higgs interactions can be generated
in the effective low-energy theory below the scale of
supersymmetry (SUSY)-breaking.  In particular, one-loop radiative
corrections, in which supersymmetric particles (squarks,
higgsinos and gauginos) propagate inside the loop,
can generate the wrong-Higgs interactions shown in Fig.~\ref{fig1}.

If the superpartners are heavy, then one can derive an effective
field theory description of the Higgs-quark Yukawa couplings
below the scale of SUSY-breaking ($M_{\rm SUSY}$), where one has
integrated out the heavy SUSY particles propagating in the loops.
The resulting effective Lagrangian is\cite{Carena:2002es}:
\beq \label{yuk}
 \mathcal{L}_{\rm yuk}^{\rm eff} = -\epsilon_{ij}(h_b+ \delta
 h_b)\psi_{b}H_d^i
\psi_Q^j  + \Delta h_b \psi_{b} H_u^{k*} \psi_{Q}^k 
 + \epsilon_{ij}(h_t + \delta h_t)\psi_{t}H_u^i\psi_Q^j
+ \Delta h_t \psi_tH_d^{k*}\psi_Q^k + {\rm h.c.}
\eeq
In the limit of $M_{\rm SUSY}\gg m_Z$,
$$
 \Delta h _b = h_b\left[\frac{2\alpha_s}{3\pi}\mu M_3
 \mathcal{I}(M_{\tilde b_1},M_{\tilde b_2},
 M_g) + \frac{h_t^2}{16\pi^2}\mu A_t \mathcal{I}
(M_{\tilde t_1}, M_{\tilde t_2}, \mu)\right]\,,
$$
where, $M_3$ is the Majorana gluino mass,
$\mu$ is the supersymmetric Higgs-mass parameter, and $\widetilde b_{1,2}$
and $\widetilde t_{1,2}$ are the mass-eigenstate bottom squarks and top
squarks, respectively.  The loop integral is given by:
$$
\mathcal{I}(a,b,c) =
\frac{a^2b^2\ln{(a^2/b^2)} +   b^2c^2\ln{(b^2/c^2)}
+ c^2a^2\ln{(c^2/a^2)}}{(a^2-b^2)(b^2-c^2)(a^2-c^2)}\,.
$$
In the limit where at least one of the arguments of $\mathcal{I}(a,b,c)$ is large,
the loop integral behaves as
$\mathcal{I}(a,b,c)\sim 1/{\rm max}(a^2,b^2,c^2)$.
Thus, in the limit where $M_3\sim \mu\sim A_t\sim M_{\tilde b}\sim M_{\tilde t}
\sim M_{\rm SUSY}\gg m_Z$, the one-loop contributions to $\Delta h_b$ do
\textit{not} decouple.

\subsection{Phenomenological consequences of the wrong-Higgs Yukawa couplings}

A consequence of the wrong-Higgs Yukawa couplings is
a $\tan\beta$-enhanced modification of certain physical observables.  To see this, we rewrite the
Higgs fields in terms of the
physical Higgs mass-eigenstates (and the Goldstone bosons):
\begin{Eqnarray}
H_d^1 &=&
\tfrac{1}{\sqrt{2}}(v\cos\beta+H^0\cos{\alpha} - h^0\sin{\alpha} +
iA^0\sin{\beta}-iG^0\cos\beta)\,,\nonumber
\\
H_u^2 &=&
\tfrac{1}{\sqrt{2}}(v\sin\beta+H^0\sin{\alpha} + h^0\cos{\alpha} +
iA^0\cos{\beta}+iG^0\sin\beta)\,, \nonumber \\
H_d^2 &=&  H^-\sin{\beta}-G^-\cos\beta\,,\nonumber \\
H_u^1 &=& H^+ \cos{\beta}+G^+\sin\beta \,,\nonumber
\end{Eqnarray}%
with $v^2\equiv v_u^2+v_d^2=(246~{\rm GeV})^2$ and $\tan\beta\equiv
v_u/v_d$.  The neutral CP-even Higgs mixing angle is denoted by $\alpha$\cite{Gunion:1984yn}.  Using \eq{yuk}, the $b$-quark mass is given by
$$
m_b = \frac{h_bv}{\sqrt{2}} \cos \beta \left(1 + \frac{\delta h_b}{h_b}
+ \frac{\Delta h_b \tan \beta}{h_b}\right)
\equiv \frac{h_bv}{\sqrt{2}}\cos \beta (1 + \Delta_b)\,,
$$
which defines the quantity $\Delta_b$.

In the limit of large $\tan\beta$ the term proportional to $\delta
h_b$ can be neglected,
in which case,
$$
\Delta_b\simeq (\Delta h_b/h_b)\tan\beta\,.
$$
Thus, $\Delta_b$ is $\tan\beta$--enhanced if $\tan\beta\gg 1$.
As previously noted, $\Delta_b$ survives in the limit of large $M_{\rm SUSY}$;
this effect does not decouple.  It can generate measurable shifts
in the decay rate for $h^0\to b\bar b$:
$$
g_{h^0 b\bar b}= -\frac{m_b}{v}\frac{\sin\alpha}{\cos\beta}
\left[1+\frac{1}{1+\Delta_b}\left(\frac{\delta h_b}{h_b}-
\Delta_b\right)\left( 1 +\cot\alpha \cot\beta \right)\right]\,.
$$
At large $\tan\beta\sim 20$---50, the value of $\Delta_b$ can be as large as 0.5 in magnitude and of
either sign, leading to a significant enhancement or suppression of the
Higgs decay rate to $b\bar b$.
If $m_{H^\pm}\gg m_Z$ (corresponding to the Higgs decoupling limit treated in section 2.3), then
\beqa
-{\sin\alpha\over\cos\beta}
&=&1-\frac{2m_Z^2\sin^2\beta\cos 2\beta}{m_{H^\pm}^2}
+\mathcal{O}\left(\frac{m_Z^4}{m_{H^\pm}^4}\right), \nonumber \\
1+\cot\alpha\cot\beta&=&-\frac{2m_Z^2}{m^2_{H^\pm}}\cos 2\beta+\mathcal{O}\left(\frac{m_Z^4}{m_{H^\pm}^4}\right)\,,
\nonumber
\eeqa
in which case $g_{h^0 b\bar b}=(m_b/v)\left[1+\mathcal{O}(m_Z^2/\mhpm^2)\right]$ approaches its SM value.

\subsection{The Decoupling Limit of the MSSM Higgs sector}

In a significant fraction of the MSSM Higgs sector parameter space,
one finds a neutral CP Higgs boson
with SM-like tree-level couplings and additional
scalar states that are somewhat heavier in mass (of order $m_{H^\pm}$),
with small mass splittings of order $m_Z^2/m^2_{H^\pm}$.  Below the scale
$m_{H^\pm}$, the effective Higgs theory coincides with that of the Standard Model (SM).

In the limit of $m_{H^\pm}\gg\mz$, the expressions for the
tree-level MSSM Higgs masses and CP-even neutral Higgs mixing angle $\alpha$ simplify\cite{Gunion:2002zf}:
\beqa
\mhl^2 &\simeq &  \mz^2\cos^2 2\beta\,,\qquad\qquad
\mhh^2 \simeq   \mha^2+\mz^2\sin^2 2\beta\,,\nonumber\\
\mhpm^2& = &  \mha^2+\mw^2\,,\qquad\qquad
\cos^2(\beta-\alpha)\simeq {\mz^4\sin^2 4\beta\over 4m_{H^\pm}^4}\,.\label{decoup}
\eeqa
From \eq{decoup}, it follows that
(i)~the two neutral heavy Higgs states and
$H^\pm$ are mass-degenerate up to corrections of ${\cal O}(\mz^2/m_{H^\pm}^2)$;
and (ii)~$\cos(\beta-\alpha)=0$ up to corrections of ${\cal O}(\mz^2/m_{H^\pm}^2)$.
This is the decoupling limit, since at energy scales below the
approximately common mass of the heavy
Higgs bosons $H^\pm$ $H^0$, $A^0$, the effective Higgs theory is
precisely that of the SM.
In general, in the limit of $\cos(\beta-\alpha)\to 0$,
all the $h^0$ couplings to SM particles approach their SM limits.

\begin{figure}[ht!]
\begin{center}
\includegraphics*[width=1.75in]{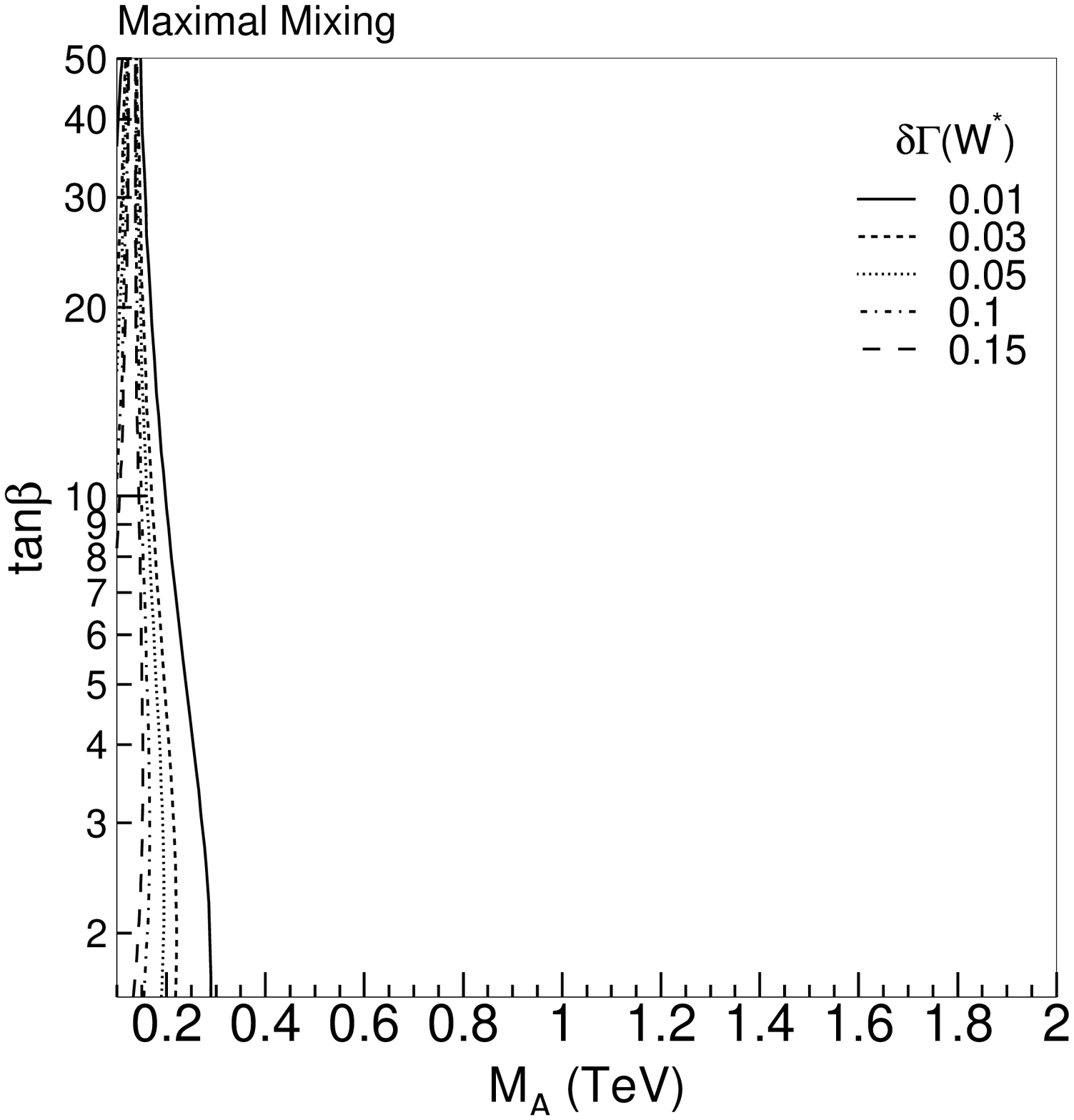}\hspace{0.05in}
\includegraphics*[width=1.75in]{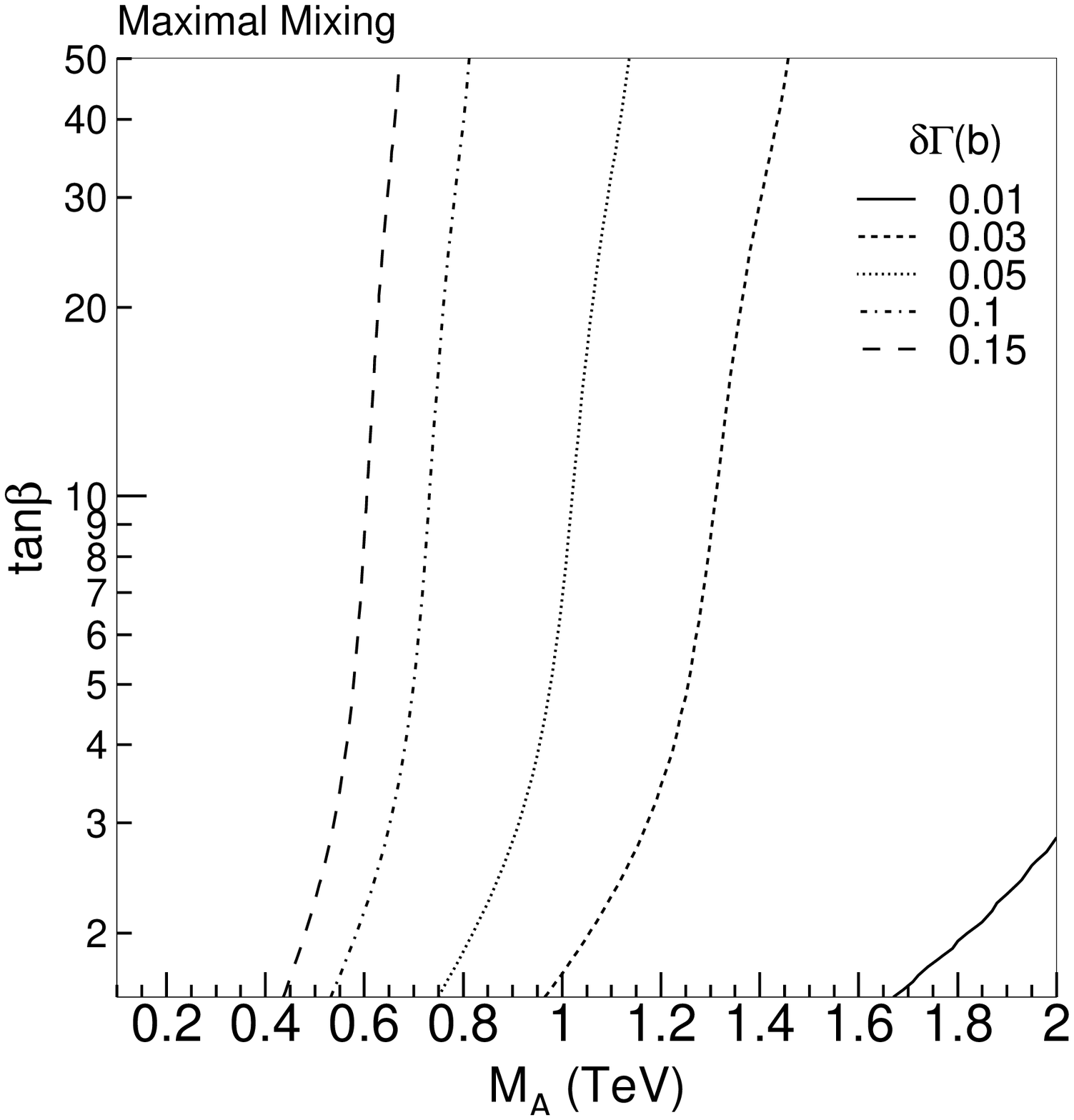}\hspace{0.05in}
\includegraphics*[width=1.75in]{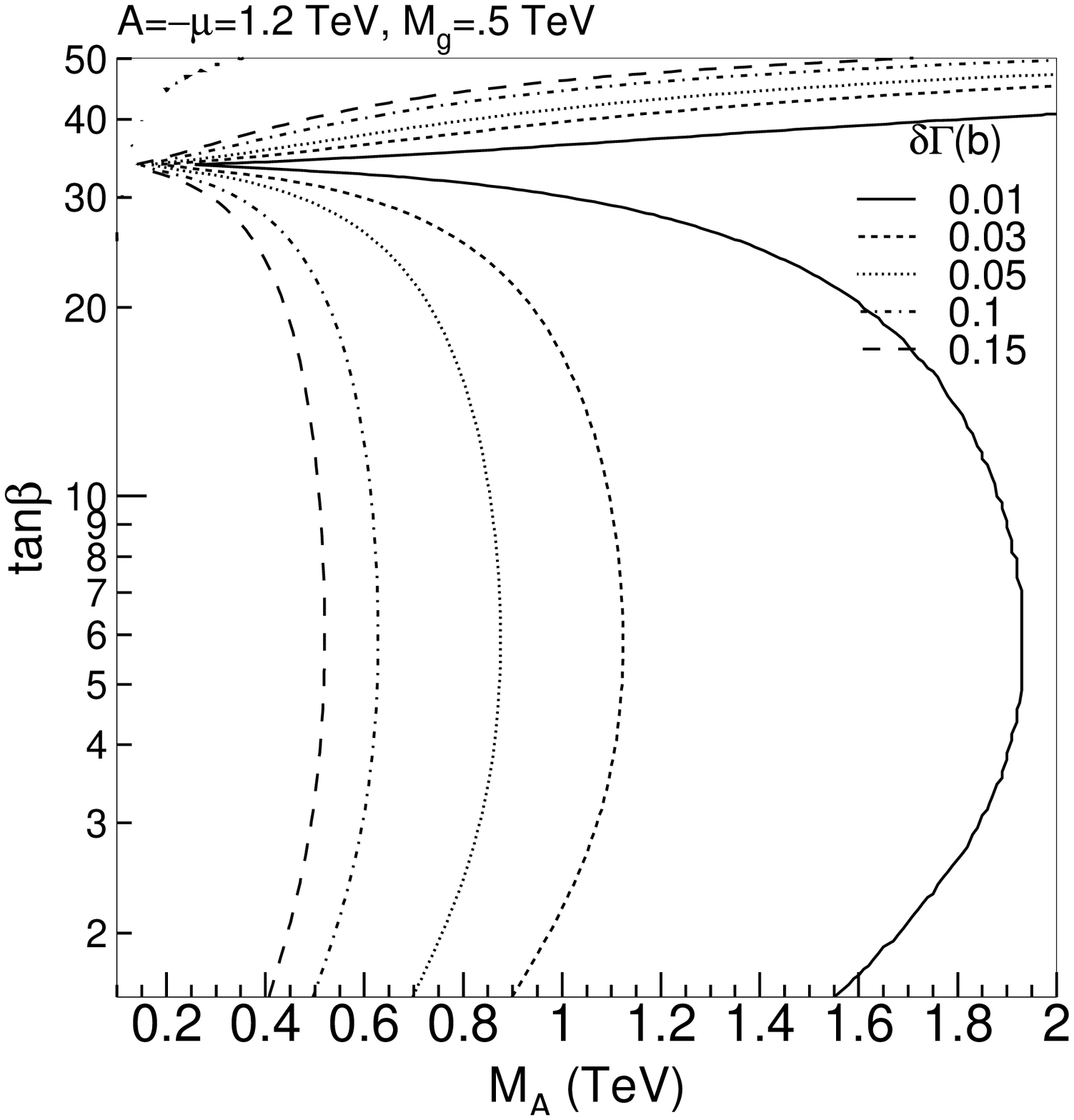}
\includegraphics*[width=1.75in]{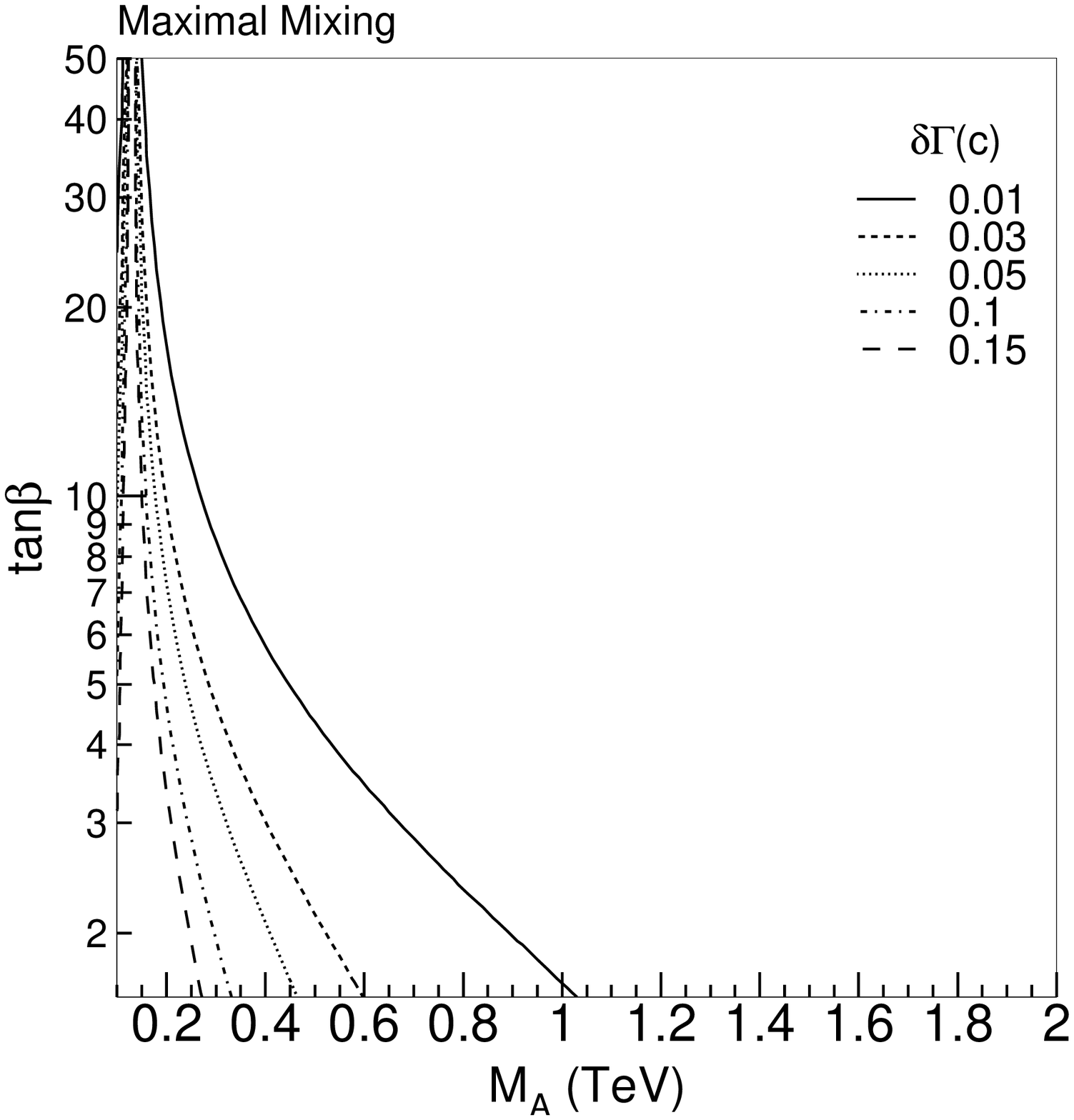}\hspace{0.05in}
\includegraphics*[width=1.75in]{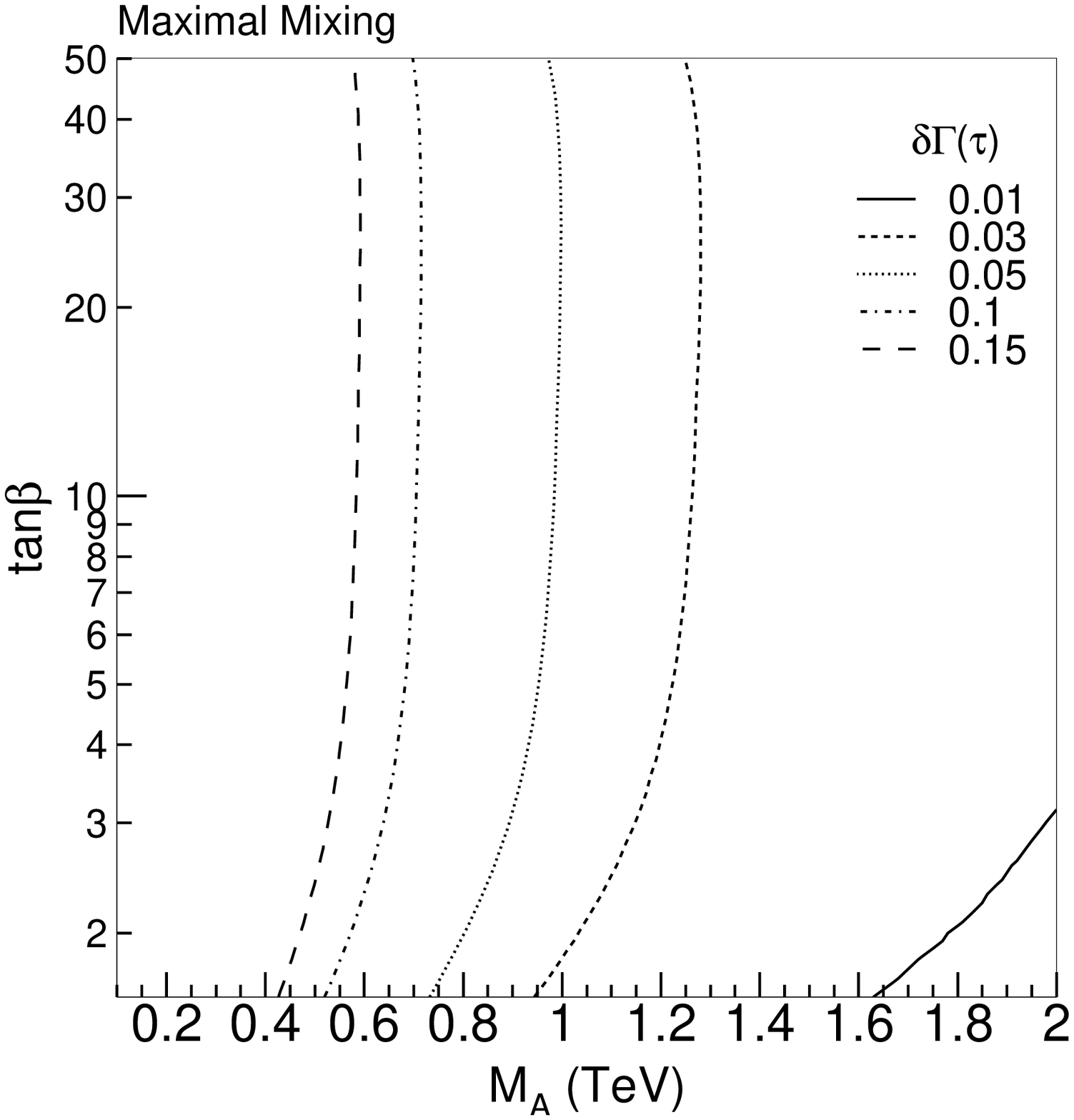}\hspace{0.05in}
\includegraphics*[width=1.75in]{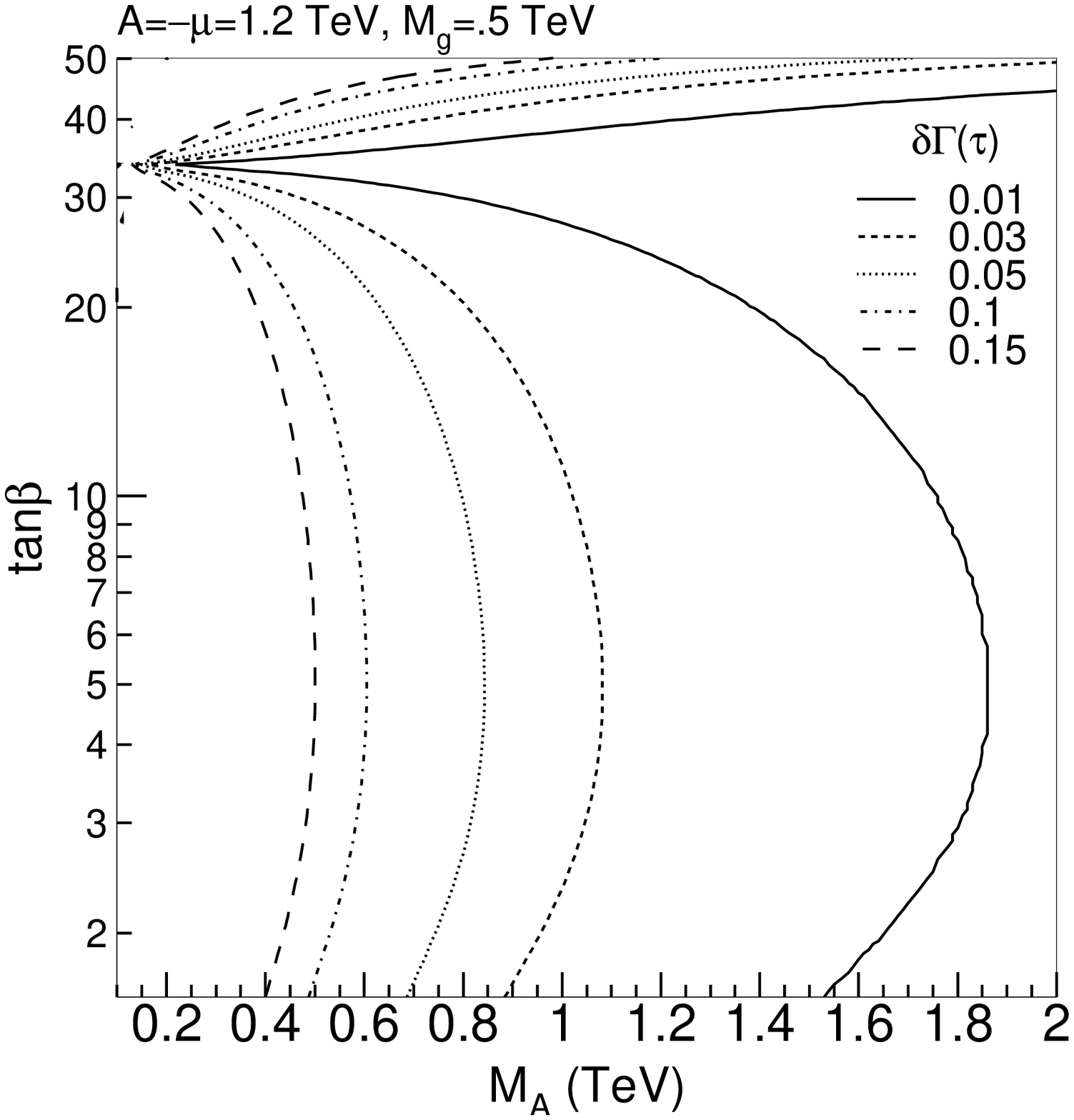}
\end{center}
\caption{\small Deviations of Higgs partial widths from
their SM values, including the leading one-loop corrections in two different MSSM scenarios, taken from Ref.\cite{Carena:2001bg}.}
\label{CHLM}
\end{figure}

These conclusion remain valid after radiative
corrections are taken into account.  Even when such effects break the CP-invariance of the tree-level MSSM Higgs
sector, the lightest neutral Higgs state possesses CP-even interactions up to small CP-violating corrections
of $\mathcal{O}(m_Z^2/\mhpm^2)$.
For example, if we keep only
the leading $\tanb$-enhanced radiative corrections, then 
in the approach to the decoupling limit\cite{Carena:2002es},
\beqa
   {g^2_{hVV}\over g^2_{\hsm VV}} & \simeq &
1-{c^2 m_Z^4\sin^2 4\beta\over 4m_{H^\pm}^4}\,, \nonumber \\[5pt]
   {g^2_{htt}\over g^2_{\hsm tt}} & \simeq & 1+{c m_Z^2\sin 4\beta
\cot\beta\over m_{H^\pm}^2}\,, \nonumber \\[5pt]
  {g^2_{hbb}\over g^2_{\hsm bb}} & \simeq & 1-{4c m_Z^2\cos 2\beta
\over m_{H^\pm}^2}\left[\sin^2\beta-{\Delta_b\over 1+\Delta_b}\right]\,,
\eeqa
where $c\equiv 1+\mathcal{O}(g^2)$ and $\Delta_b\equiv
\tan\beta\times \mathcal{O}(g^2)$, with $g$ a generic gauge or
Yukawa coupling.  The quantities $c$ and $\Delta_b$ depend on the
MSSM spectrum. The approach to decoupling is fastest for the $h^0$
couplings to vector boson pairs and slowest for the couplings to
down-type quarks\cite{Carena:2001bg}, as shown in Fig.~\ref{CHLM}.

\section{The general Two-Higgs-Doublet Model (2HDM)}

Consider the most general renormalizable 2HDM potential,

\beqa
\mathcal{V}&=&m_{11}^2\Phi_1^\dagger\Phi_1+m_{22}^2\Phi_2^\dagger\Phi_2
-[m_{12}^2\Phi_1^\dagger\Phi_2+{\rm h.c.}]\nn\\[6pt]
&&\qquad\qquad +\half\lambda_1(\Phi_1^\dagger\Phi_1)^2 
+\half\lambda_2(\Phi_2^\dagger\Phi_2)^2
+\lambda_3(\Phi_1^\dagger\Phi_1)(\Phi_2^\dagger\Phi_2)
+\lambda_4(\Phi_1^\dagger\Phi_2)(\Phi_2^\dagger\Phi_1)\nn\\[6pt]
&&\qquad\qquad 
+\left\{\half\lambda_5(\Phi_1^\dagger\Phi_2)^2
+\big[\lambda_6(\Phi_1^\dagger\Phi_1)
+\lambda_7(\Phi_2^\dagger\Phi_2)\big]
\Phi_1^\dagger\Phi_2+{\rm h.c.}\right\}\,. \label{thdm}
\eeqa 
In a general 2HDM, $\Phi_1$ and $\Phi_2$ are indistinguishable weak-doublet, hypercharge-one fields.
A basis change consists of a global U(2) transformation $\Phi_a\to
U_{a\bbar}\Phi_b$ (and  $\Phi_\abar^\dagger=
\Phi_\bbar^\dagger U^\dagger_{b\abar}$).  Note that the electroweak gauge-covariant kinetic energy
terms of the scalar fields are invariant with respect to U(2), whereas the scalar potential
squared-masses and couplings change under U(2) transformations and thus
are \textit{basis-dependent} quantities.
Physical quantities that can be measured in the laboratory must be
basis-independent.  Thus, any model-independent experimental study of 2HDM phenomena
must employ basis-independent methods for analyzing data associated with 2HDM physics.

The most general 2HDM defined by \eq{thdm} generically contains large
tree-level Higgs-mediated FCNCs and
CP-violating effects, which are inconsistent with present experimental
data over a large range of the 2HDM parameter space.  This can be rectified by either
(i)~fine-tuning of 2HDM parameters to reduce the size of the FCNC and CP-violating
effects below the experimentally allowed limits; or
(ii)~imposing additional symmetries (discrete and/or continuous)
on the Higgs Lagrangian to eliminate tree-level Higgs-mediated FCNCs and
CP-violation.  The latter can distinguish between $\Phi_1$ and $\Phi_2$,
in which case a choice of basis acquires physical significance.

However, any experimental analysis that relies on a specific symmetry and corresponding basis
choice cannot be used to interpret the data in terms of a more general 2HDM.
In contrast, basis-independent methods can be employed to perform model-independent
2HDM analyses.  Moreover, such an analysis could in principle  be used to
experimentally identify and distinguish among possible approximate or exact symmetries.
Finally, if these symmetries are badly broken or nonexistent, then the basis-independent
methods are the only viable framework for 2HDM studies.

\subsection{The basis-independent formalism}

The scalar potential can be rewritten in U(2)-covariant notation\cite{branco,Davidson:2005cw}:
$$
\mathcal{V}=Y_{a\bbar}\Phi_{\abar}^\dagger\Phi_b+\half
Z_{a\bbar
  c\dbar}(\Phi_{\abar}^\dagger\Phi_b)(\Phi_{\cbar}^\dagger\Phi_d)\,,\qquad a,b,c,d=1,2\,,
$$
where $Z_{a\bbar c\dbar}=Z_{c\dbar a\bbar}$ and hermiticity implies
$Y_{a \bbar}= (Y_{b \abar})^\ast$ and
$Z_{a\bbar c\dbar}= (Z_{b\abar d\cbar})^\ast$.
The barred indices help keep track of which indices transform with
$U$ and which transform with~$U^\dagger$. For example,
$Y_{a\bbar}\to U_{a\cbar}Y_{c\dbar}U^\dagger_{d\bbar}$
and $Z_{a\bbar c\dbar}\to U_{a\ebar}U^\dagger_{f\bbar}U_{c\gbar}
U^\dagger_{h\dbar} Z_{e\fbar g\hb}$ under a U(2) transformation.

The vacuum expectation values of the two Higgs fields
can be parametrized as
\beq \label{emvev} \langle
\Phi_a \rangle={\frac{v}{\sqrt{2}}} \left(
\begin{array}{c} 0\\ \widehat v_a \end{array}\right)\,,\qquad
{\rm with}\qquad
\widehat v_a \equiv e^{i\eta}\left(
\begin{array}{c} \cos\beta\,\\ e^{i\xi}\sin\beta \end{array}\right)
\,,
\eeq
where $v=246$~GeV, $0\leq\beta\leq\half\pi$ and $\eta$ is arbitrary.  It is convenient to define
$V_{a\bbar}\equiv \hat v_a \hat v_{\bbar}^*$, which is an hermitian matrix
with orthonormal eigenvectors $\hat v_b$ and
$\widehat w_b \equiv
\widehat v_\cbar^{\,\ast}\,\epsilon_{cb}$.  (In our conventions,
$\epsilon_{12}=-\epsilon_{21}=1$ and $\epsilon_{11}=\epsilon_{22}=0$.)
In particular, note that $\hat v_{\bbar}^* \hat w_b=0$.
Under a U(2) transformation, 
\beq \label{chidef}
\hat v_a\to U_{a\bbar}\hat v_b\,,
\qquad\quad \widehat w_a\to e^{-i\chi}\,U_{a\bbar\,} \widehat w_b\,,
\eeq
where ${\rm det}~U\equiv e^{i\chi}$ is a pure phase.
That is, $\widehat w_a$ is a pseudo-vector with respect to U(2).
One can use $\widehat w_a$
to construct a proper second-rank tensor,
$W_{a\bbar}\equiv \hat w_a \hat w_{\bbar}^*
\equiv\delta_{a\bbar}-V_{a\bbar}$.  Moreover $\tan\beta\equiv |\hat v_2/\hat v_1|$ is
basis-dependent.  Hence, $\tan\beta$ is \textit{not} in general a physical parameter.

All 2HDM observables must be invariant under a basis transformation $\Phi_a\to U_{a\bbar}\Phi_b$.
Invariants under a basis transformation $\Phi_a\to U_{a\bbar}\Phi_b$
are obtained by combining the
tensor quantities $Y$, $Z$, $\widehat{v}$ and $\widehat{w}$.
All physical observables can be expressed in invariant combinations of real invariants and complex 
pseudo-invariants:\footnote{Pseudo-invariants are useful because
one can always combine two such quantities to create an invariant.}
\beqa
Y_1 &\equiv& \Tr(YV)\,,\qquad\qquad\qquad\quad\! Y_2 \equiv
\Tr(YW)\,,\qquad\qquad\qquad\,\, Y_3 \equiv
Y_{a\bbar}\,\widehat v_\abar^\ast\, \widehat w_b\,,
\nonumber \\
Z_1 &\equiv& Z_{a\bbar c\dbar}\,V_{b\abar}V_{d\cbar}\,,\qquad\qquad\quad\!
Z_2 \equiv Z_{a\bbar c\dbar}\,W_{b\abar}W_{d\cbar}\,,\qquad\qquad
Z_3 \equiv Z_{a\bbar
c\dbar}\,V_{b\abar}W_{d\cbar}\,,\nonumber \\
Z_4 & \equiv &
Z_{a\bbar c\dbar}\,V_{b\cbar}W_{d\abar}
\qquad\qquad\quad\,
Z_5 \equiv Z_{a\bbar c\dbar}\,\widehat v_\abar^\ast\, \widehat w_b\,
\widehat v_\cbar^\ast\, \widehat w_d\,,\nonumber \\
Z_6 &\equiv&
Z_{a\bbar c\dbar}\,\widehat v_\abar^\ast\,\widehat v_b\,
\widehat v_\cbar^\ast\, \widehat w_d\,,\qquad\quad\,
Z_7 \equiv Z_{a\bbar c\dbar}\,\widehat v_\abar^\ast\, \widehat w_b\,
\widehat w_\cbar^\ast\,\widehat w_d\,. \label{inv}
\eeqa %
For example, the charged Higgs boson mass (which is clearly a physical observable) is given by
$$
m_{H^\pm}^2=Y_{2}+\half Z_3 v^2\,.
$$

The complex pseudo-invariants listed in \eq{inv} transform as
$$ [Y_3, Z_6, Z_7]\to e^{-i\chi}[Y_3, Z_6, Z_7] \quad{\rm and}\quad
Z_5\to  e^{-2i\chi} Z_5\,,$$
where $\chi$ is defined below \eq{chidef}.  As an example,
the scalar potential minimum conditions,
$$
Y_1=-\half Z_1 v^2\quad \text{and}\quad Y_3=-\half Z_6 v^2\,, 
$$
are covariant conditions with respect to U(2) transformations.

\subsection{The Higgs basis and Higgs mass eigenstate basis}

The three physical neutral Higgs boson mass-eigenstates, denoted by $h_1$, $h_2$ and $h_3$,
are determined by diagonalizing a $3\times 3$ real symmetric squared-mass
matrix that is defined in a basis
where only one of the two
neutral Higgs fields has a non-zero vacuum expectation value---the so-called
\textit{Higgs basis}\cite{branco,haberoneil}.
The diagonalizing matrix is a $3\times 3$
real orthogonal matrix that depends on three angles:
$\theta_{12}$, $\theta_{13}$ and~$\theta_{23}$.
Under a U(2) transformation\cite{haberoneil},
$$
\theta_{12}\,,\, \theta_{13}~{\hbox{\text{are invariant, and}}}\,\,
e^{i\theta_{23}}\to e^{-i\chi}\, e^{i\theta_{23}}\,.
$$
One can express the mass eigenstate
neutral Higgs field directly in terms of the original shifted neutral
fields,  $\overline\Phi_a\lsup{0}\equiv \Phi_a^0-v\widehat
v_a/\sqrt{2}$:
\beq \label{hk}
h_k=\frac{1}{\sqrt{2}}\left[\overline\Phi_{\abar}\lsup{0\,\dagger}
(q_{k1} \widehat v_a+q_{k2} e^{-i\theta_{23}}\widehat w_a)
+(q^*_{k1}\widehat v^*_{\abar}+q^*_{k2}
e^{i\theta_{23}}\widehat w^*_{\abar}) \overline\Phi_a\lsup{0}\right]\,, 
\eeq
for $k=1,\ldots,4$, where $h_4\equiv G^0$ is the neutral Goldstone field.

\begin{wraptable}{r}{0.38\columnwidth}
\vskip -0.1in
\begin{tabular}{|c||c|c|}\hline
$k $ &\phaa $q_{k1}\phaa $ & \phaa $q_{k2} \phaa $ \\
\hline
$1$ & $c_{12} c_{13}$ & $-s_{12}-ic_{12}s_{13}$ \\
$2$ & $s_{12} c_{13}$ & $c_{12}-is_{12}s_{13}$ \\
$3$ & $s_{13}$ & $ic_{13}$ \\
$4$ & $i$ & $0$ \\ \hline
\end{tabular}
\caption{\small The $q_{k\ell}$ are
functions of the angles $\theta_{12}$ and
$\theta_{13}$, where
$c_{ij}\equiv\cos\theta_{ij}$ and $s_{ij}\equiv\sin\theta_{ij}$.}\label{tabq}
\vskip -0.2in
\end{wraptable}

The \textit{invariant}
quantities $q_{k\ell}$ are listed in Table~\ref{tabq}.
Since $\widehat w_a e^{-i\theta_{23}}$ is a \textit{proper}
U(2)-vector, we see that the neutral mass-eigenstate fields are indeed
invariant under basis transformations.  Likewise, $H^+$
and its charge conjugate (as defined below) are U(2)-invariant fields.
Inverting \eq{hk} yields:
\beq
\Phi_a=\left(\begin{array}{c}G^+\widehat v_a+ e^{-i\theta_{23}} H^+ \widehat w_a\\[6pt]
\displaystyle \frac{v}{\sqrt{2}}\widehat
v_a+\frac{1}{\sqrt{2}}\sum_{k=1}^4 \left(q_{k1}\widehat
v_a+q_{k2}e^{-i\theta_{23}}\widehat w_a\right)h_k
\end{array}\right).\nn
\eeq

\subsection{Basis-independent form for the Higgs couplings of the 2HDM}

The Higgs boson interactions of the 2HDM can be expressed in terms of the basis-independent
$q_{k\ell}$ defined in Table \ref{tabq}. The cubic and quartic vector-scalar
couplings were obtained in Ref.\cite{haberoneil}.  A slightly more convenient form for these couplings, obtained
in Appendix A of Ref.\cite{Haber:2010bw}, is reproduced below:\footnote{The results of this section differ
from  Refs.\cite{haberoneil} and \cite{Haber:2010bw} by a rephasing of the charged Higgs fields so that the $H^\pm$ are
the invariant fields as defined below \eq{hk}.}
\beqa
\mathscr{L}_{VVH}&=&\left(gm_W W_\mu^+W^{\mu\,-}+\frac{g}{2c_W}
m_Z Z_\mu Z^\mu\right)q_{k1} h_k \nonumber \\[5pt]
&&
+em_WA^\mu(W_\mu^+G^-+W_\mu^-G^+)
-gm_Zs_W^2 Z^\mu(W_\mu^+G^-+W_\mu^-G^+)
\,, \label{VVH}
\\[8pt]
\mathscr{L}_{VVHH}&=&\left[\quarter g^2  W_\mu^+W^{\mu\,-}
+\frac{g^2}{8c_W^2}Z_\mu Z^\mu\right]h_k h_k+\biggl\{\half g^2 W_\mu^+ W^{\mu\,-}+e^2A_\mu A^\mu
\nonumber \\[5pt]
&&\qquad\qquad \quad +\frac{g^2}{c_W^2}\left(\half - s_W^2\right)^2Z_\mu Z^\mu
+\frac{2ge}{c_W}\left(\half -s_W^2\right)A_\mu Z^\mu\biggr\}(G^+G^-+H^+H^-)
\nonumber \\[5pt]
&& +\biggl\{\left(\half eg A^\mu W_\mu^+ -\frac{g^2s_W^2}{2c_W}Z^\mu W_\mu^+\right)
(q_{k1}G^-+q_{k2}H^-)h_k +{\rm h.c.}\biggr\}
\,,\label{VVHH}\\[8pt]
\mathscr{L}_{VHH}&=&\frac{g}{4c_W}\,\epsilon_{jk\ell}q_{\ell 1}
Z^\mu h_j\ddel_\mu h_k
-\half g\biggl\{iW_\mu^+\left[q_{k1} G^-\ddel\lsup{\,\mu} h_k+
q_{k2}H^-\ddel\lsup{\,\mu} h_k\right]
+{\rm h.c.}\biggr\}\nonumber \\[5pt]
&&+\left[ieA^\mu+\frac{ig}{c_W}\left(\half -s_W^2\right)
Z^\mu\right](G^+\ddel_\mu G^-+H^+\ddel_\mu H^-)\,,\label{VHH} \\[8pt]
\mathscr{L}_{VG}&=&\left[\frac{g^2}{4} W_\mu^+W^{\mu\,-}
+\frac{g^2}{8c_W^2}Z_\mu Z^\mu\right]G^0 G^0+\half g\left(W_\mu^+G^-\ddel\lsup{\,\mu}G^0+W_\mu^-G^+\ddel\lsup{\,\mu}G^0
\right)\,\nonumber\\[5pt]
&&+\biggl\{\frac{ieg}{2} A^\mu W_\mu^+ G^- G^0
-\frac{ig^2s_W^2}{2c_W}Z^\mu W_\mu^+
G^-G^0 +{\rm h.c.}\!\biggr\}+\frac{g}{2c_W} q_{k1} Z^\mu G^0\ddel_\mu h_k\,,\label{VG}
\eeqa
where repeated indices $j,k=1,2,3$ are summed over.
Terms quadratic in the scalar fields that contain one or two neutral Goldstone fields are exhibited in \eq{VG}.

Likewise, a basis-independent form for the cubic and quartic scalar self-interactions
has been obtained in Ref.\cite{haberoneil}.  The cubic couplings are given by:  
\beqa 
\mathcal{V}_3&=&\half v\, h_j h_k h_\ell
\biggl[q_{j1}q^*_{k1}\Re(q_{\ell 1}) Z_1
+q_{j2}q^*_{k2}\,\Re(q_{\ell 1})(Z_3+Z_4) +
\Re(q^*_{j1} q_{k2}q_{\ell 2}Z_5\,
e^{-2i\theta_{23}}) \nonumber \\
&&\qquad\qquad\qquad\quad
+\Re\left([2q_{j1}+q^*_{j1}]q^*_{k1}q_{\ell 2}Z_6\,e^{-i\theta_{23}}\right)
+\Re (q_{j2}^*q_{k2}q_{\ell 2}Z_7\,e^{-i\theta_{23}})
\biggr]\nonumber \\
&& \hspace{-0.2in} +v\,h_k
G^+G^-\biggl[\Re(q_{k1})Z_1+\Re(q_{k2}\,Z_6 e^{-i\theta_{23}})\biggr]
+v\,h_k H^+H^-\biggl[\Re(q_{k1})Z_3+\Re(q_{k2}\,Z_7 e^{-i\theta_{23}})\biggr]
\nonumber \\
&& \hspace{-0.2in} +\half v \,h_k\biggl\{G^-H^+
\left[q^*_{k2} Z_4
+q_{k2}\,Z_5 e^{-2i\theta_{23}}+2\Re(q_{k1})\,Z_6 e^{-i\theta_{23}} \right]+{\rm h.c.}\biggr\}\,,\nn
\eeqa
where repeated indices $j,k,\ell=1,2,3,4$ are summed over.
The neutral Goldstone fields are implicitly included by denoting $h_4\equiv G^0$.
In particular, $\Re(q_{k1})=q_{k1}$ for $k=1,2,3$, whereas $\Re(q_{41})=0$.
With the same conventions as above, the quartic scalar couplings are given by:
\beqa
\mathcal{V}_4&=&\eighth h_j h_k h_l h_m
\biggl[q_{j1}q_{k1}q^*_{\ell 1}q^*_{m1}Z_1
+q_{j2}q_{k2}q^*_{\ell 2}q^*_{m2}Z_2
+2q_{j1}q^*_{k1}q_{\ell 2}q^*_{m2}(Z_3+Z_4)\nonumber \\[5pt]
&&
+2\Re(q^*_{j1}q^*_{k1}q_{\ell 2}q_{m2}Z_5\,e^{-2i\theta_{23}})
+4\Re(q_{j1}q^*_{k1}q^*_{\ell 1}q_{m2}Z_6\,e^{-i\theta_{23}})
+4\Re(q^*_{j1}q_{k2}q_{\ell
  2}q^*_{m2}Z_7\,e^{-i\theta_{23}})\biggr]\nonumber \\[5pt]
&&  +\half h_j h_k G^+ G^-\biggl[q_{j1}q^*_{k1} Z_1 + q_{j2}q^*_{k2}Z_3
+2\Re(q_{j1}q_{k2}Z_6\,e^{-i\theta_{23}})\biggr]
 \nonumber \\[5pt]
&&  +\half h_j h_k H^+ H^-\biggl[q_{j2}q^*_{k2} Z_2 + q_{j1}q^*_{k1}Z_3
+2\Re(q_{j1}q_{k2}Z_7\,e^{-i\theta_{23}})\biggr] \nonumber \\
&& +\half h_j h_k\biggl\{G^- H^+ \left[q_{j1}q^*_{k2}Z_4
+ q^*_{j1}q_{k2}Z_5\,e^{-2i\theta_{23}}+q_{j1}q^*_{k1}Z_6
\,e^{-i\theta_{23}}+q_{j2}q^*_{k2}Z_7\,e^{-i\theta_{23}}\right]+{\rm h.c.}
\biggr\} \nonumber \\[5pt]
&&
 +\half Z_1 G^+ G^- G^+ G^- +\half Z_2 H^+H^- H^+ H^-
+ (Z_3+Z_4)G^+ G^- H^+ H^- 
\nonumber \\[5pt]
&&
+\biggl\{\half Z_5 e^{-2i\theta_{23}} H^+H^+G^-G^- + Z_6 e^{-i\theta_{23}} G^+G^- H^+ G^-
 +Z_7 e^{-i\theta_{23}} H^+H^- H^+ G^- +{\rm h.c.}\biggr\}
\,,\nn
\eeqa
summing over $j$, $k$, $\ell$, $m=1,2,3,4$. 

The Higgs couplings to quarks and leptons are determined by the Yukawa Lagrangian.
In terms of the quark mass-eigenstate fields, 
$$
-\mathscr{L}_{\rm Y}=\anti U_L \Phi_{\abar}^{0\,*}h^U_a \ur -\anti
D_L K^\dagger\Phi_{\abar}^- h^U_a\ur
+\anti U_L K\Phi_a^+h^{D\,\dagger}_{\abar} \dr
+\anti D_L\Phi_a^0 h^{D\,\dagger}_{\abar}\dr+{\rm h.c.}\,,
$$
where
$\wtil\Phi_{\abar}\equiv (\wtil\Phi^0_{\abar}\,,\,\wtil\Phi^-_{\abar})=i\sigma_2\Phi_{\abar}^*$ and
$K$ is the CKM mixing matrix.  The $h_a^{U,D}$
are $3\times 3$ Yukawa coupling matrices.  
We can construct invariant and pseudo-invariant matrix Yukawa couplings:
$$
\kappa^{Q}\equiv \widehat v^*_\abar h^{Q}_a\quad {\rm and} \quad
\rho^{Q}\equiv \widehat w^*_\abar h^{Q}_a\,,
$$
where $Q=U$ or $D$.  Inverting these equations yields $h^Q_a=\kappa^Q\widehat
v_a+\rho^Q\widehat w_a$.
Under a U(2) transformation, $\kappa^Q$ is
invariant, whereas $\rho^Q\to e^{i\chi}\rho^Q$.

By construction, $\kappa^U$ and $\kappa^D$ are proportional to the
(real non-negative) diagonal quark mass matrices $M_U$ and $M_D$,
respectively, whereas the matrices $\rho^U$ and $\rho^D$ are independent complex
$3\times 3$ matrices.  In particular,
$$
M_U=\frac{v}{\sqrt{2}}\kappa^U={\rm diag}(m_u\,,\,m_c\,,\,m_t)\,,\qquad
M_D=\frac{v}{\sqrt{2}}\kappa^{D\,\dagger}={\rm
diag}(m_d\,,\,m_s\,,\,m_b) \,.
$$

The final form for the Yukawa couplings of the mass-eigenstate Higgs
bosons and the Goldstone bosons to the quarks, is given by\cite{haberoneil}: 
\beqa
 &&  -\mathscr{L}_Y = \frac{1}{v}\overline D
\biggl\{M_D (q_{k1} P_R + q^*_{k1} P_L)+\frac{v}{\sqrt{2}}
\left[q_{k2}\,[e^{i\theta_{23}}\rho^D]^\dagger P_R+
q^*_{k2}\,e^{i\theta_{23}}\rho^D P_L\right]\biggr\}Dh_k \nonumber \\
&&\qquad\quad  +\frac{1}{v}\overline U \biggl\{M_U (q_{k1}
P_L + q^*_{k1} P_R)+\frac{v}{\sqrt{2}}
\left[q^*_{k2}\,e^{i\theta_{23}}\rho^U P_R+
q_{k2}\,[e^{i\theta_{23}}\rho^U]^\dagger P_L\right]\biggr\}U h_k
\nonumber \\
&&\quad  +\biggl\{\overline U\left[K[e^{i\theta_{23}}\rho^D]^\dagger
P_R-[e^{i\theta_{23}}\rho^U]^\dagger KP_L\right] DH^+ +\frac{\sqrt{2}}{v}\,\overline
U\left[K\mdd P_R-\mud KP_L\right] DG^+ +{\rm
h.c.}\biggr\}\,,\nonumber
\eeqa
where $P_{L,R}=\half(1\mp\gamma\ls{5})$.
Indeed, the Higgs-fermion Yukawa couplings
$\mathscr{L}_Y$ depends only on invariant quantities:
the $3\times 3$ matrices $M_Q$ and $\rho^Q e^{i\theta_{23}}$ and the
invariant angles $\theta_{12}$ and~$\theta_{13}$.  Note that the unphysical parameter $\tan\beta$
does \textit{not} appear.

The couplings of the neutral Higgs bosons to quark pairs are
generically flavor-nondiagonal and CP-violating, since the
$q_{k2}$ and the matrices $e^{i\theta_{23}}\rho^Q$ are
not generally either pure real or pure imaginary.

\subsection{Symmetries of the Higgs-fermion interactions}

A general 2HDM exhibits CP-violating neutral Higgs boson couplings to fermions
and tree-level FCNCs mediated by neutral Higgs boson exchange.  Both effects
can be removed by imposing an appropriate symmetry.
Once again, a basis-independent formulation of
such symmetries is useful (as these could be determined in principle from experimental data).

The conditions for a tree-level CP-conserving neutral Higgs--quark interactions are given by\cite{Haber:2010bw}:\footnote{CP symmetry cannot be exact due to
the unremovable phase in the CKM matrix
that enters via the charged current interactions mediated by either $W^\pm$, $H^\pm$
or $G^\pm$ exchange.}
$$Z_5(\rho^Q)^2\,,\, Z_6\rho^Q~~\text{and}~~Z_7\rho^Q~~\text{are real matrices ($Q=U$, $D$).}
$$
Type-I and Type-II Higgs-quark interactions are defined as follows\cite{types1and2}:
\begin{description}
 \item{Type I:} \,$\epsilon_{\abar\bbar}h^D_a h^U_b=
\epsilon_{ab}h^{D\,\dagger}_{\abar}h^{U\,\dagger}_{\bbar}=0$, \qquad\quad i.e., $h_2^U=h_2^D=0$ in some basis;
\item{Type II:} $\delta_{a\bbar}\,h^{D\,\dagger}_{\abar}h^U_b = 0$, \,\qquad\qquad\qquad\qquad i.e.,
$h_1^U=h_2^D=0$ in some basis,
\end{description}
which can be implemented with a $\mathbb{Z}_2$ symmetry (with appropriate choices
for the transformations of the scalar and fermion fields), or with supersymmetry.

Invariant expressions for the Type-I and Type-II conditions are given by\cite{Davidson:2005cw,haberoneil}:
$$
\text{Type I:}~~~ \kappa^D\rho^U=\rho^D\kappa^U\,,\qquad\quad
\text{Type II:}~~~\kappa^D\kappa^U+\rho^{D\,\dagger}\rho^U=0\,,
$$
where in both cases, $\rho^Q\propto \kappa^Q=\sqrt{2}M_Q/v$ (for $Q=U$, $D$).  Hence,
in both cases there are no off-diagonal neutral Higgs--quark couplings.  The generalization to 
Higgs--lepton couplings is straightforward.

The existence of a special basis (up to an overall rephasing of the Higgs fields) in which $h_2^U=h_2^D=0$ (Type-I) or
$h_1^U=h_2^D=0$ (Type-II) promotes $\tan\beta$ to a physical parameter, where $\tan\beta$
is defined to be the
magnitude of the ratio of the neutral Higgs vacuum expectation values in the special basis.

For example, suppose we define the invariant parameters
$$
 \tan\beta_D\equiv
 \frac{v}{3\sqrt{2}}\,\bigl|\Tr\left(\rho^D M_D^{-1}\right)\bigr|\,,
 \qquad
 \tan\beta_U\equiv\frac{\sqrt{2}}{3v}\,\bigl|\Tr\left([\rho^U]^{-1} M_U\right)\bigr|\,.
$$
Then, in a Type-II model these two quantities coincide
and one can identify the physical parameter $\tan\beta=\tan\beta_D=\tan\beta_U$.
Thus, in a model-independent analysis, measuring $\tan\beta_D$ and
$\tan\beta_U$ can shed light on the symmetries of the Higgs boson--quark Yukawa couplings.

\section{The decoupling limit in the general 2HDM}

In the decoupling limit, one of the two Higgs doublets of the 2HDM
receives a very large mass and is therefore decoupled from the
theory.  This is achieved when
$Y_2\gg v^2$ and $|Z_i|\lsim\mathcal{O}(1)$ [for all~$i$].
The effective low energy theory is then a one-Higgs-doublet
model, corresponding to the SM Higgs sector.

If we order the neutral scalar masses according to $m_1< m_{2,3}$
and define the Higgs mixing angles accordingly, then
the basis-invariant conditions for the decoupling limit are\cite{haberoneil}:
\beq
|s_{12}|\lsim\mathcal{O}\left(\frac{v^2}{m_2^2}\right)\ll 1\,,
\qquad\quad  |s_{13}|\lsim\mathcal{O}\left(\frac{v^2}{m_3^2}\right)\ll 1\,,
\qquad\quad
\Im(Z_5\,e^{-2i\theta_{23}})\lsim\mathcal{O}
\left(\frac{v^2}{m_3^2}\right)\ll 1\,.\nn
\eeq

In the
decoupling limit,  $m_1\ll m_2, m_3, m_{H^\pm}$.
In particular, the properties of $h_1$ coincide with the SM-like Higgs boson
with $m_1^2=Z_1 v^2$ up to corrections
of $\mathcal{O}(v^4/m^2_{2,3})$, whereas
$m_2\simeq m_3\simeq \mhpm$ with squared mass splittings of
$\mathcal{O}(v^2)$.
In contrast,
far from the decoupling limit, one typically finds that \textit{all} Higgs
bosons have a similar mass of $\mathcal{O}(v)$ and \textit{none} are SM-like.

In the exact decoupling limit [$s_{12}=s_{13}=\Im(Z_5\,e^{-2i\theta_{23}})=0$],
the interactions of
$h_1$ are precisely those of the SM Higgs boson.
In particular,
the corresponding interactions of $h_1$
are CP-conserving and flavor-diagonal.
In the approach to decoupling limit of a general 2HDM, the
CP-violating and flavor-changing neutral Higgs couplings of
the SM-like Higgs state $h_1$
are suppressed by factors of $\mathcal{O}(v^2/m^2_{2,3})$.
In contrast,
the corresponding interactions of the heavy neutral Higgs bosons ($h_2$
and $h_3$) and the charged Higgs bosons ($\hpm$)
can exhibit both CP-violating and flavor non-diagonal couplings
(proportional to the~$\rho^Q$).

The decoupling limit is a generic feature of extended Higgs sectors\cite{Haber:1989xc}.\footnote{However,
if some terms of the Higgs potential are absent, it is possible that no
decoupling limit may exist.  In this case, the only way to have very large Higgs
masses is to have large Higgs self-couplings.}
Thus, the initial observation of a SM-like Higgs boson (at LHC) does not rule out the possibility
of an extended Higgs sector in the decoupling regime.  In particular, only a
precision Higgs program can reveal small deviations from the
decoupling limit, which would signal the existence of a new heavy mass scale associated with
the heavier Higgs states or a manifestation of other new physics beyond the Standard Model.

\section{Conclusions and lessons for future work}

If a Higgs boson is discovered at the LHC, with properties that approximate those of
the SM Higgs boson, then one must establish a precision Higgs program to determine the
structure of the scalar dynamics responsible for electroweak symmetry breaking.
A precision Higgs program at the LHC can at best measure the Higgs couplings to
gauge bosons and fermions with an accuracy of 10--20\%\cite{Duhrssen:2004cv}.
A precision Higgs program at ILC and/or CLIC can achieve significantly improved
accuracies, in some cases by an order of magnitude\cite{precisionILC,precisionCLIC}.

Precision measurements of the properties of a
SM-like Higgs may reveal small deviations, which can indicate
the presence of a non-minimal Higgs sector and/or new physics beyond
the Standard Model (characterized by a new mass scale that lies above the scale
of electroweak symmetry-breaking).  With sufficient precision, the measurement
of Higgs couplings can be sensitive
to a mass scale of new physics that lies beyond the reach of the collider [cf.~Fig.~\ref{CHLM}].


Basis-independent methods provide
a powerful technique for studying the theoretical structure of
the two-Higgs doublet model.  These methods provide insight
into the conditions for CP-conservation (and violation\cite{Gunion:2005ja}), as well
as other exact or approximate symmetries that 
govern the 2HDM dynamics and can distinguish between the
two Higgs doublet fields.  If nature suggests an elementary scalar sector
with the structure of the 2HDM, then basis-independent techniques
will be essential for performing a model-independent analysis
to determine the $\rho^Q$ ($Q=U,D$) and eventually the $Z_i$.
The collider tools for such an analysis have yet to be fully developed.


\section*{Acknowledgments}

I am especially grateful to Sacha Davidson and Deva O'Neil who collaborated
with me in developing the basis-independent framework for the 2HDM presented
in this talk.  I also acknowledge my other co-authors, Marcela Carena, Jack Gunion, Heather Logan, 
John Mason and Steve Mrenna, for their invaluable contributions. 
This work is supported
in part by the U.S. Department of Energy, under grant
number DE-FG02-04ER41268.



\begin{footnotesize}


\end{footnotesize}


\end{document}